\documentclass[fleqn,10pt]{wlscirep}
\usepackage{graphicx}
\usepackage{epstopdf}
\title{Machine-learned approximations to Density Functional Theory Hamiltonians}

\author[1,*]{Ganesh Hegde}
\author[1]{R. Chris Bowen}
\affil[1]{Advanced Logic Lab, Samsung Semiconductor Inc., Austin, TX 78754, USA}
\affil[*]{ganesh.h@ssi.samsung.com}

\keywords{Machine Learning, Hamiltonians, Tight Binding}

\begin{abstract}
Large scale Density Functional Theory (DFT) based electronic structure calculations are highly time consuming and scale poorly with system size. While semi-empirical approximations to DFT result in a reduction in computational time versus \textit{ab initio} DFT, creating such approximations involves significant manual intervention and is highly inefficient for high-throughput electronic structure screening calculations. In this letter, we propose the use of machine-learning for prediction of DFT Hamiltonians. Using suitable representations of atomic neighborhoods and Kernel Ridge Regression, we show that an accurate and transferable prediction of DFT Hamiltonians for a variety of material environments can be achieved. Electronic structure properties such as ballistic transmission and band structure computed using predicted Hamiltonians compare accurately with their DFT counterparts. The method is independent of the specifics of the DFT basis or material system used and can easily be automated and scaled for predicting Hamiltonians of any material system of interest. 
\end{abstract}
\begin{document}

\flushbottom
\maketitle
%
%

\section*{Introduction}

Density Functional Theory (DFT) based electronic structure calculations are a cornerstone of modern computational material science. Despite their popularity and usefulness, DFT methods scale rather poorly with system size. Even the fastest DFT techniques, such as the so-called O(N) approaches based on Local Orbitals spend most of their time in forming the Hamiltonian and solving self-consistently for the ground state electron eigenstates \cite{bowler2012methods}. Consequently, while it scales well for small systems such as crystalline solids with small unit cells, scaling the computation of DFT electronic structure for large nanostructures with limited periodicity becomes a challenging task.

Semi-empirical Tight Binding (TB) approximations to DFT perform rather admirably in this regard, reducing the time complexity in forming Hamiltonians by treating Hamiltonian elements as parameters fitted to desired physical properties such as effective mass, band gaps and so on. Typically, TB techniques also simplify the DFT Hamiltonian to near-neighbor (NN) interactions between atoms. This results in extremely sparse Hamiltonians that can be diagonalized very efficiently. Several flavors of these approximations exist, generally optimized to meet the needs of the problem at hand \cite{goringe1997tight}. 

While TB calculations have proved immensely useful in modeling the electronic structures of semiconductor and metallic nanostructures, some fundamental problems persist in their use as DFT Hamiltonian approximations. Firstly, while TB Hamiltonians are computationally efficient to solve for, what is often unaccounted for is the time required to create an accurate, physically transparent and transferable TB model for a material system. In our own experience \cite{klimeck2000si,hegde2014environment}, developing meaningful, physical TB parameter sets has taken time ranging from a few weeks to a few years. Secondly, developing truly transferable, physically valid TB models applicable across materials, geometries and boundary conditions is challenging, owing to physical differences in systems. 

There are a variety of reasons for the aforementioned shortcomings. One, it is often the case that a physically transparent model created for a particular material type (say semiconductors with short range, covalent bonding) does not generalize to other material systems (for instance, metals). Secondly, significant portions of the TB parameter-set creation process involve manual intervention. From the functional form of the model used to fit Hamiltonian elements to the choice and tuning of the fitting process, the TB parameterization process often resembles art than an exact science. 

This is clearly an undesirable state of affairs, especially in time-sensitive industrial applications where a number of material systems, geometries and boundary conditions need to be evaluated rapidly to test for suitability for an application. To address these issues, we propose a Machine Learning (ML) based method to predict DFT Hamiltonians. In contrast to existing semi-empirical approximations to DFT, our method does not fit parametric models to DFT Hamiltonians or derived electronic structure quantities such as effective masses and band gaps. We instead map atomic environments to real-space Hamiltonian predictions using non-parametric interpolation-based machine learning techniques such as Kernel Ridge Regression (KRR) \cite{friedman2001elements}. The time-consuming direct computation of Hamiltonian integrals and ground state eigenstates in DFT is completely bypassed in favor of simple matrix multiplication reducing computational cost considerably. Through rigorous criteria for selection of algorithm hyper-parameters, the method can be automated and transferability to new material systems can be controlled. The method is independent of the specifics of the material system and depends only on the reference DFT calculations. 
    
The paper is organized as follows. The methods section describes how the problem of predicting DFT Hamiltonians can be mapped on to a ML framework. The results section shows the application of the method proposed in this work to two material systems - a arbitrarily strained Cu system and a arbitrarily strained C (diamond) system. The Cu case study assesses the applicability of the present technique to a relatively simple Cu-DFT system consisting only of rotation invariant $s$-orbitals centered on Cu atoms, while the C case study extends the technique to more realistic systems involving $s$ and $p$ orbitals and their interactions. A comparison of the advantages and limitations of the method to existing semi-empirical approximations to DFT is included in the discussion section.

\section*{Methods}
In recent times, several advances have been made in applying a variety of machine learning techniques to computational material science. These include (but are not limited to) the mapping of spatial atomic data to predict total energies \cite{bartok2010gaussian, rupp2012fast, behler2007generalized}, prediction of DFT functionals \cite{snyder2012finding} and direct computation of electronic properties \cite{schutt2014represent} through machine learning. We are, however, unaware of any attempt to apply ML to the problem of DFT Hamiltonian prediction. 

Applying ML to any problem presupposes the existence of data and a unknown, non-trivial map from inputs to outputs. In supervised ML, the dataset consists of inputs/variables (henceforth referred to as 'features') and the challenge is to find a map from features to pre-labeled/desired/reference outputs (targets). In our case, the dataset consists of inputs in the form of sets of atomic positions while the targets consist of DFT Hamiltonian elements between orbitals centered on individual atoms. The reference data is obtained by time consuming self-consistent DFT calculations, but the aim is that once a map is obtained from input to reference data, the map is sufficiently accurate and transferable to cases not explicitly included in the dataset.

Real-space DFT Hamiltonians consist of intra-atomic (between orbitals situated on the same atom) and inter-atomic (between orbitals situated on different atoms) matrix elements. Most real-space DFT codes allow direct access to these elements upon the convergence of a ground state DFT calculation. In case of plane-wave DFT codes, a projection of plane-wave based Hamiltonian can be made on to a real space basis. The rest of this paper assumes a real-space Hamiltonian. These real-space DFT Hamiltonian matrix elements then constitute our reference data. 

Broadly speaking, intra-atomic and inter-atomic (real-space) DFT Hamiltonian elements depend on
\begin{enumerate}
\item The position and charges of the interacting atoms. 
\item The positions and charges of atoms that are in the neighborhood of interacting atoms.
\item The basis functions (orbitals) centered on each atom.
\end{enumerate} 
1) and 2) above define a 3D-radially and angularly varying potential (upon DFT calculation convergence) that is highly non-trivial for the simplest of systems that go beyond a single atom. Factor 3) above is usually fully known \textit{a priori}. From this enumeration, one can surmise that obtaining useful features required for ML requires a faithful representation of the positions and charges of interacting atoms and their atomic neighborhood. 

A possible starting point in forming features for ML is the collection of atomic positions and charges (atomic numbers).  However, operations such as rotation, translation and permutation of atom numbers in an atomic dataset change their positions and their relative ordering completely. In general, Hamiltonian elements are are invariant to translation and permutation, but change upon a rotation reference frame (higher spherical harmonics are not rotation invariant). It is therefore important that the features formed reflect these properties. The atomic positions and charges must therefore be transformed to reflect the interactions between orbitals and the atomic environment responsible in forming DFT Hamiltonian interactions.

Seminal work published in the past 10 years has realized the difficult task of invariant representation of atomic neighborhoods possible to the point where these approaches can readily be used in ML for a variety of computational material science applications. These include the Bispectrum formalism of Bartok et al. \cite{bartok2010gaussian}, the Coulomb Matrix formalism (see for instance, Rupp et al. \cite{rupp2012fast} and Montavon et al. \cite{montavon2012learning}), the Symmetry Function formalism of Behler et al. \cite{behler2007generalized}, the Partial Radial Distribution Function (PRDF) formalism of Sch{\"u}tt et al. \cite{schutt2014represent}, the Bag-of-Bonds (BoB) formalism of Hansen et al.\cite{hansen2015machine} and the Fourier Series of RDF formalism of von Lilienfeld et al.\cite{von2015fourier}, to name a few. In this work, we have used the bispectrum formalism of Bartok et al. to represent the neighborhood of an atom. In this formalism, the 3D radial and angular environment of each atom is converted into vector of bispectrum components of the local neighbor density projected on to a basis of hyperspherical harmonics in four dimensions. While a detailed discussion of the the formalism is beyond the scope of this work, we refer the interested reader to the literature on the topic \cite{bartok2013representing, bartok2010gaussian, thompson2015spectral, bartok2015gaussian}. 

\begin{figure}[!htbp]
	\centering
		\includegraphics[width=3in,height=2.5in]{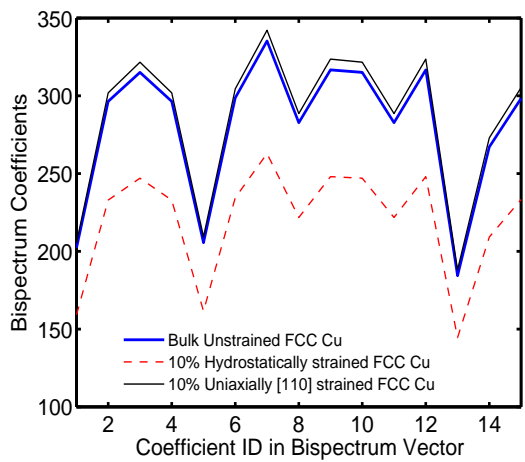}
	\caption{Bispectrum measure for a Cu atom in 3 different FCC environments - bulk unstrained, 10\% hydrostatic strain and 10\% uniaxial strain along [100]. The Bispectrum is a unique measure for the atom in the different environments and is therefore an excellent candidate for the creation of ML features from atomic spatial data.}
	\label{fig:Figure1_Bispectrum}
\end{figure}

For the purpose of this work, it suffices to understand the bispectrum as a unique, vectorized representation of an atoms local environment, so that it can be used as a basis for creating features suitable for machine learning. Figure \ref{fig:Figure1_Bispectrum} is a simple example of how the bispectrum distinguishes between different strained environments that a Face Centered Cubic (FCC) Cu atom is exposed to. The various components of the bispectrum vector can reflect subtle variations in neighbor environment, creating a unique map between features and Hamiltonian elements. Our choice of bispectrum is motivated by its rotation, translation and permutation invariant representation of the \textit{atomic environment} in a convenient vectorized form, with an ability to tune the dimensionality of the bispectrum vector by changing the number of angular momentum channels. We hasten to add that it should not be construed as distinctly advantageous by the reader and alternate representations of the atoms environment that share these features may be considered in place of the bispectrum without affecting the applicability of the present method.

\begin{figure}
	\centering
		\includegraphics[width=\textwidth]{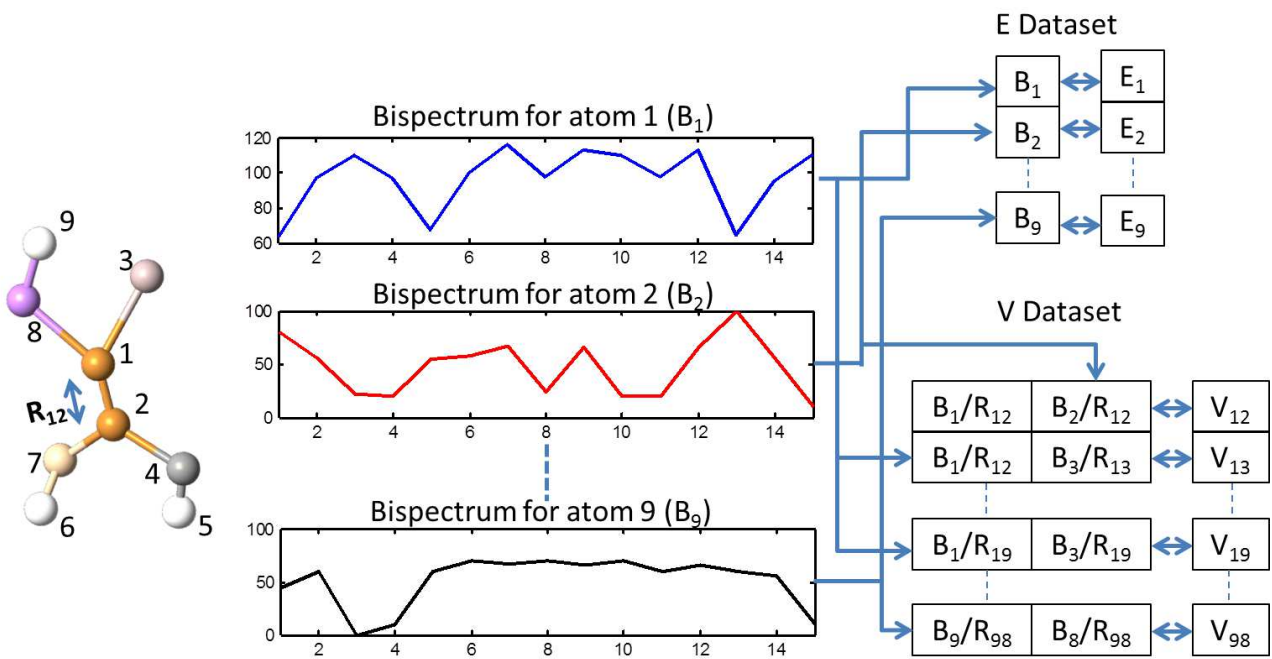}
	\caption{Schematic representation of the reference dataset creation process for the Cu-case study. First the atomic structure data in the form of atom types, charges and positions are converted to corresponding bispectra. The bispectra of different atoms are then combined to form features for learning. For intra-atomic Hamlitonian elements (E), the bispectra of individual atoms constitute the feature vector. For inter-atomic Hamiltonian elements (V), the bispectra are divided by the inter-atomic distances (R) to form the feature vector. The feature vectors along with their corresponding reference output data constitute the reference dataset.}
	\label{fig:Fig5_BisSchema}
\end{figure}

$s$-orbitals and their interactions with other s-orbitals are rotation, translation and permutation invariant. For the Cu-case study where a single $s$-orbital centered on Cu atoms forms the basis set, our input features can then be simply formed from the bispectra of the atoms taking part in the interaction without explicit information about the reference frame or its rotation. For intra-atomic elements in the Cu-case study, the feature vector we use is simply the bispectrum of the respective atoms. For inter-atomic Hamiltonian elements in this case study, a candidate feature vector is the bispectrum of the two atoms taking part in the interaction. Since 3D periodic material systems contain an infinite number of atoms having the same neighborhood, we divide the bispectra of the two atoms taking part in the Hamiltonian interaction by their inter-atomic distance and append the interatomic distance to form the feature vector. We found that adding some measure of the interatomic distance explicitly in the feature vector in addition to its implicit inclusion by (modifying the bispectrum) systematically improved predictive performance of our method. For instance we found that the RMSE of the inter-atomic interactions reduced by approximately 2 meV with the inclusion of interatomic distance. We note, however, that we did not systematically investigate the performance of a distance metric being explicitly included in the feature vector versus, say, inverse distance or other powers of interatomic distance.  Figure \ref{fig:Fig5_BisSchema} in the methods section shows the feature formation process for the Cu-case study in greater detail.

For the C (diamond) - case study, the simplest DFT basis set that captures electronic interactions accurately is one comprised of a single $s$ orbital and 4 $p$ orbitals. While interactions between $s$ orbitals are rotation invariant as mentioned previously, interactions between dissimilar $s$ and $p$ orbitals and between dissimilar $p$ orbitals on different atoms in general are not rotation invariant. In addition to capturing the atomic environment, which may yet be invariant to rotations, one therefore needs to capture angular information as well. For interatomic interactions, the direction cosines ($l$, $m$, $n$) of the bond vector (with respect to the Cartesian axes/laboratory reference frame) between two atoms taking part in the Hamiltonian interaction were appended to the feature vector used in the Cu-case study. Introduction of this angular information simply captures the necessary angular information for interatomic elements. 

The orbitals involved in intra-atomic interactions belong to the same atom, while the potential contribution to the Hamiltonian element is from this atom and other neighboring atoms. Rotation of the reference frame or addition of atoms in the neighborhood leaves the relative angular information of the orbitals situated on the same atom unchanged i.e. still zero. The angular information of the neighborhood, however, changes and with this rotation, the off-diagonal intra-atomic Hamiltonian elements change as well. Unlike inter-atomic interactions, however, we cannot solve this problem by simply appending neighbor direction cosines, since the interaction captured is between two orbitals on the same atom. This is addressed by appealing to the physics of the problem at hand. The neighbors that exert the most influence on the intra-atomic interaction are the nearest neighbors. We therefore sum the direction cosines with respect to each Cartesian axis from each neighbor and append this to the bispectrum. For instance, a C atom in diamond-like configurations has 4 nearest neighbors, with direction cosines $l_{1}$, $l_{2}$, $l_{3}$ and $l_{4}$ with respect to the $x$ Cartesian axis. For unstrained diamond, the sum of these direction cosines is zero as is the sum of direction cosines with respect to $y$ and $z$ axes. This captures very well the actual DFT calculation for the intra-atomic block in diamond, where off-diagonal intra-atomic interactions are zero. When symmetry is broken by straining uni-axially for instance, the sum of direction cosines becomes non-zero . This is reflected in the corresponding Hamiltonian where the off-diagonal intra-atomic elements become non-zero. This scheme outlined above is one among a number of such schemes that can be devised to incorporate angular interactions.

In the Slater-Koster scheme used in classical Tight-Binding \cite{podolskiy2004compact}, introducing angular information in the form of direction cosines can be shown to represent the angular part of all inter-atomic Hamiltonian interactions, while canonical interatomic interactions between orbitals  - $V_{ss\sigma}, V_{sp\sigma}, V_{pp\pi}, V_{pp\sigma}$ - can be obtained by aligning the bond-axis to the laboratory/Cartesian $z$-axis reference. Generally speaking, however, interatomic interactions go beyond the simple-two center scheme and the Slater Koster scheme applied fails to account for more complicated 'three-center' interactions. This is true for intra-atomic interactions as well, where one needs to go beyond the two-center approximation to capture non-zero off-diagonal intra-atomic Hamiltonian elements. Since the aim in machine learning is typically to let the data 'find' a reliable map upon supplying features sufficient to represent the outputs meaningfully, we have simply chosen to incorporate angular information in the form of direction cosines and their sums instead of imposing a model on the underlying interactions. 

Having formed the features thus, the non-linear map from the features to the Hamiltonian elements can be learned using a ML technique called Kernel Ridge Regression (KRR). In KRR, the predicted output for a given feature vector is expressed as a combination of similarity measures between the feature vector (for which an output is to be predicted) and all other feature vectors in the dataset.
Using the notation from Rupp\cite{rupp2015machine}, for a feature vector $\tilde{x}$, the predicted output $f(\tilde{x})$ is given as
\begin{equation}\label{eqn:fx}
f(\tilde{x}) = \sum_{i=1}^{n} \alpha_{i}k(x_i,\tilde{x})
\end{equation}

Here, $n$ is the number of training examples, $k(x_i,\tilde{x})$ is the kernel measure between feature vectors $x_i$ and $\tilde{x}$ and $\alpha_{i}$ is the corresponding weight in the sum. The kernel measure $k(x_i,\tilde{x})$ allows a comparison of similarity (or conversely, distance) between feature vectors \cite{friedman2001elements}. We use a Gaussian kernel defined as follows

\begin{equation}\label{eqn:kernel}
k(x_i,\tilde{x}) = \exp(-\gamma{\lVert x_i - \tilde{x} \rVert_{2}}^{2})
\end{equation}

$\gamma$ is an algorithm hyper-parameter that decides the extent (in feature space) to which the kernel operates. The second term in the exponent is the squared 2-norm of the difference vector between the two feature vectors $x_i$ and $\tilde{x}$. The similarity kernel returns a value of 1 when the feature vectors are identical, indicating maximum similarity. By changing the value of $\gamma$ one can tune the extent to which the kernel returns similarity values. For $\gamma$ values close to zero, only those feature vectors that are closely matched will return a similarity measure approaching 1, whereas feature vectors that are dissimilar will return values approaching zero. For larger values of $\gamma$, a finite similarity value is returned by the kernel for feature vectors that are unequal.

Equation \ref{eqn:fx} above thus interpolates between existing feature vectors in the dataset and predicts a value based on the similarity of the candidate feature vector to those in the dataset. The weights $\alpha$ are learned by computing the similarity measure between all feature vectors in the training dataset.
\begin{equation}\label{eqn:w}
\alpha = (K+\lambda I)^{-1}y
\end{equation}
where the kernel matrix $K$ is obtained by computing similarity measures using equation \ref{eqn:kernel} between each feature vector in the training dataset, $y$ is the vector of training reference data. $\lambda$ is a regularization hyper-parameter that controls the smoothness of the interpolator and prevents over-fitting. $I$ is the identity matrix.

Equations \ref{eqn:fx}, \ref{eqn:kernel} and \ref{eqn:w} together constitute the equations defining the KRR learning algorithm. Given a training dataset $X$ (with $n$ feature vectors $x_i$ and i from 1 through $n$) and $y$ (with reference outputs $y_i$ for each feature vector in the training dataset), these equations can be used to train the set of weights $\alpha$, obtain a similarity measure $K$ for the dataset, and predict the output value for a candidate feature vector $\tilde{x}$ for a given value of the algorithm hyper-parameters $\lambda$ and $\gamma$.

Having shown how a DFT Hamiltonian problem can be mapped on to a machine learning problem, we now summarize the requirements from  a procedural perspective. First, $n$ reference self-consistent DFT calculations on the material system of interest are performed. For each DFT calculation, DFT Hamiltonian elements are extracted to form the reference outputs $y$. For each reference output $y_i$, a feature vector $x_i$ is extracted by converting the atomic positions, charges and neighborhoods of the atoms involved in the formation of matrix elements $y_i$ to a corresponding invariant bispectrum measure using the aforementioned procedure. The reference dataset $(X,y)$ thus formed is partitioned into training $(X_{train},y_{train})$ and test datasets $(X_{test},y_{test})$. Figure \ref{fig:Fig5_BisSchema} is a schematic representation of the creation of the procedure used to create the reference datasets.

The training dataset is used to learn the weights $w$ that map feature vectors $X_{train}$ to outputs $y_{train}$ using equation \ref{eqn:kernel} and \ref{eqn:w} above for the set of hyper-parameters $\lambda$ and $\gamma$. Equation \ref{eqn:fx} can then be used to predict Hamiltonian elements. The quality of the fit can be assessed by comparing predicted values $f(\tilde{X}_{test})$ for features in the test dataset with the reference values $y_{test}$.

The selection of algorithm hyperparameters $\lambda$ and $\gamma$ is done by creating a grid of $\lambda$ and $\gamma$ values and evaluating a goodness-of-fit metric for each combination of hyperparameters in the grid. A number of statistical metrics such as the $R^2$ score, the root mean square (RMS) error, the absolute error can be used. The hyperparameter combination that results in the best goodness-of-fit metric can be used for training. Using this technique, algorithm hyper-parameters can be rigorously and automatically chosen instead of basing the selection on trial-and-error. The dataset generated initially can be augmented continuously and the algorithm weights and hyper-parameters can easily be re-learned for maximum transferability. In addition, a significant portion of the prediction procedure - learning weights, hyper-parameters and predicting Hamiltonian elements - can be parallelized. For large datasets, efficient algorithms such as gradient descent can be used instead of the linear equations above to optimize algorithm weights \cite{friedman2001elements}. The accuracy and transferability of the technique can be further improved by techniques such as $K$-fold cross validation, where the training and validation is repeatedly performed on smaller partitions of the dataset \cite{rupp2015machine,friedman2001elements} and testing is performed on the test dataset that does not enter the fitting procedure.  We used $K$=5-fold cross validation for all the results in this work.

\begin{figure}[!htbp]
	\centering
		\includegraphics[width=3in,height=2.5in]{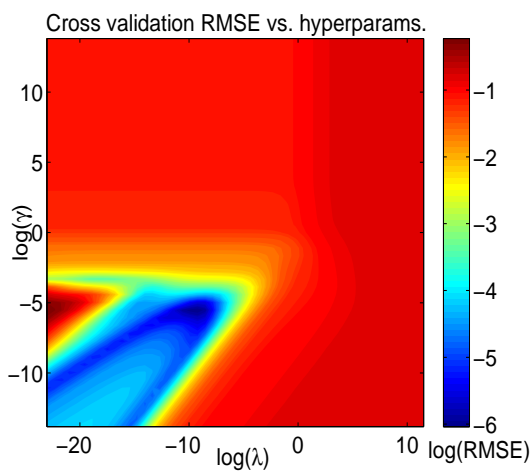}
	\caption{Grid of algorithm hyper-parameters $\gamma$ and $\lambda$ versus Root Mean Square Error (RMSE) scores for the V dataset for the C-diamond case study. A large number of possible combinations of hyper-parameters result in a similar range of RMSE and can be used for training weight $\alpha$. We used the combination that provided the lowest RMSE across both case studies.}
	\label{fig:Figure4_AGS}
\end{figure}

Our ML technique of choice in this work was KRR with Gaussian kernels. We mention, however, that our choice of kernel is not unique. For instance, Laplacian kernels applied to the study of molecular energies \cite{rupp2015machine} appear to have improved performance in comparison to Gaussian kernels. Alternative machine learning algorithms such as Gaussian Process Regression (GPR) \cite{rasmussen2006gaussian} and Support Vector Regression (SVR) \cite{smola2004tutorial} can perform the same task with similar accuracy and ease of implementation \cite{abu2012learning}.

Since the feature vectors created incorporate charge and distance information, the method is independent of the specifics of the material system used and the same algorithm can be applied without change to a variety of material systems. The material system and DFT specifics-agnostic nature of the method makes it suitable for high-throughput electronic-structure screening calculations where a number of machines can be learned rapidly for multiple materials and Hamiltonian predictions can be made to account for changes in electronic structure versus orientation, periodicity, disorder and so on.

\section*{Results}
As mentioned previously, we demonstrate the soundness of the method with the help of two case studies. The first involves the prediction of Hamiltonian elements in hydrostatically and uniaxially strained FCC Cu. For simplicity and ease of demonstration, we a use a compensated single $s$-orbital basis for Cu \cite{hegde2015feasibility} for DFT calculations. The second case study, involving multi-orbital systems involved the uniaxial and hydrostatically strained diamond phase of Carbon.
\subsection{Single $s$-orbital strained Cu system Cu}
For the single $s$-orbital basis used in the Cu-case study, the DFT Hamiltonian consists of only two types of elements
\begin{enumerate}
\item $E_s$ - The intra-atomic or on-site energy element for the Cu $s$ orbital. We henceforth refer to this element as E. E captures the effect of the external environment (potential created by the neighborhood of atoms) on a given Cu atom.
\item $V_{ss}$ - The inter-atomic or off-site energy element that captures the strength of the interaction between $s$-orbitals situated on interacting atoms mediated by the potential created by the neighborhood of interacting atoms. We henceforth refer to this element as V.
\end{enumerate}

To create the reference dataset, we performed self-consistent DFT calculations for a total of 200 uniaxially and hydrostatically strained bulk FCC Cu systems. This results in a reference/labeled dataset with 27000 intra and inter-atomic Hamiltonian elements for the strained Cu system. The details of the DFT calculations for Cu such as the basis set, the compensation charge, the DFT functional, the $k$-point grid density are outlined in work we recently published \cite{hegde2015feasibility}. The specifics of the DFT technique used are unimportant for this work, since the ML technique is independent of the type and details of DFT calculation. One only needs to ensure that a consistent $k$-point grid density and DFT functional is used across all reference calculations so that the calculations are consistent across boundary conditions.

\begin{figure}[!htbp]
	\centering
		\includegraphics[width=\textwidth]{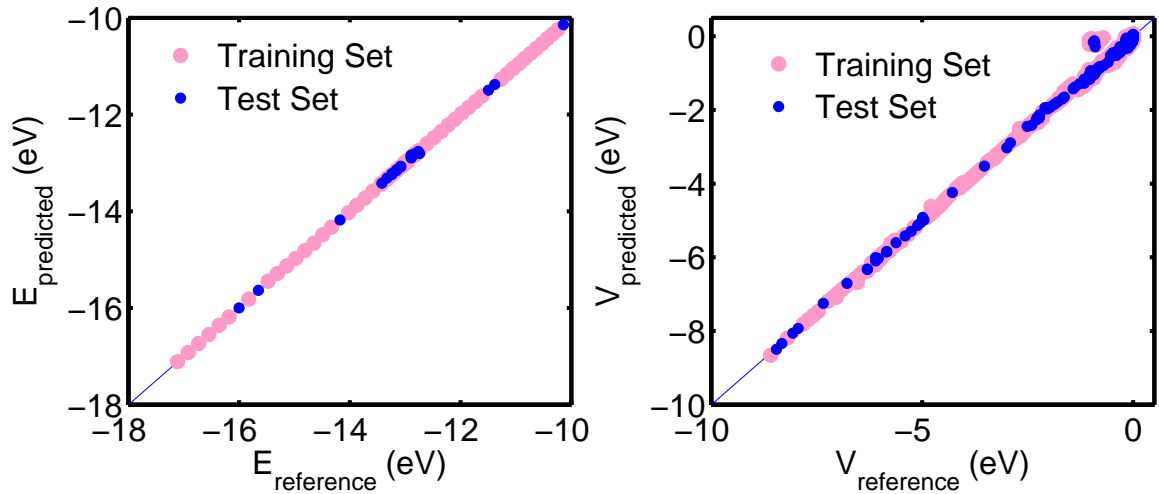}
	\caption{Predicted versus reference intra-atomic interaction energies (left) and inter-atomic interaction energies (right) for the single $s$-orbital Cu system studied. The $x=y$ line is included as an indicator for the quality of fit - when predictions match reference/desired data perfectly, they lie on this line.}
	\label{fig:Figure2_EV}
\end{figure}

From the converged DFT calculations, we extract E values for each atom in the calculation and V values for each (atom, neighbor) combination to form reference output data $y_E$ and $y_V$. The E reference data set for the Cu case-study consisted of 500 intra-atomic Hamiltonian elements. The feature vector for E prediction for an atom is a row vector consisting of 69 bispectrum coefficients for that atom. The reference feature matrix $X_E$ is thus a $N$$\times$69 matrix where $N$ = 500 is the total number of atoms across all reference calculations for the Cu case study. The V reference dataset consisted of 26500 Hamiltonian elements. The V dataset is significantly larger than the E dataset since it captures interactions between each of the atoms in the E dataset and \textit{all} of its neighbors. The spatial extent of the $s$-orbital used in the DFT calculations is up to approximately 8$\AA$. Consequently, each atom in the E dataset has (averaging over all strain configurations), approximately 133 neighbors for Cu. The feature vector for V prediction consists of 69 bispectrum coefficients for the first and second atoms each divided by the interatomic distance. The interatomic distance is also appended at the end of this vector for a total dimensionality of 139 for V features in the dataset. The reference feature matrix $X_v$ is thus a 26500$\times$139 matrix. 

Machines for E and V prediction were then trained using the procedure outlined previously. We use a randomized training:test split of 80:20 percent for E and V training and prediction. We use the Root Mean Square Error (RMSE) between reference and predicted test data as a goodness-of-fit metric to test the quality of our predictions.


\begin{figure}[!htbp]
	\centering
		\includegraphics[width=\textwidth]{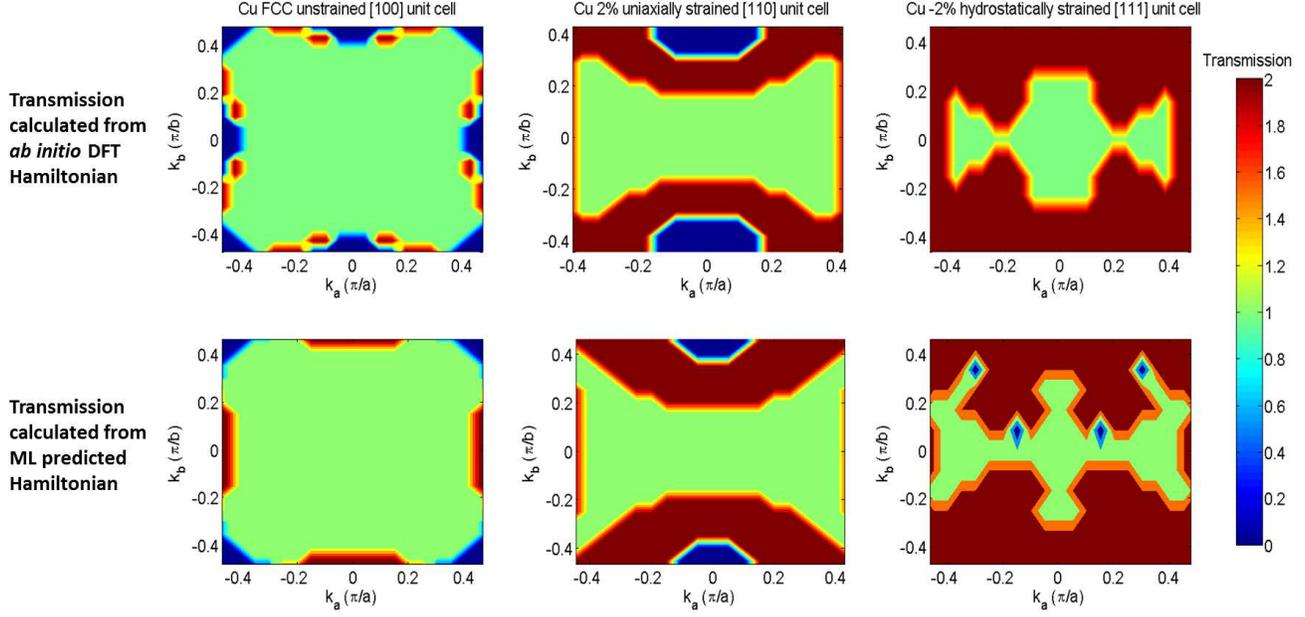}
	\caption{Examples of ballistic Transmission computed using the Hamiltonians predicted by ML versus the same calculation done \textit{ab initio} in DFT for the single $s$-orbital Cu system studied. From left to right these are 1) transverse-momentum resolved transmission in a bulk unstrained FCC Cu unit cell oriented along [100] 2) transverse momentum resolved transmission in a [110] 2$\%$ tensile unaixially strained Cu unit cell oriented along [110].  3) Transverse momentum resolved transmission in a 2$\%$ compressive hydrostatically strained Cu unit cell oriented along [111]. }
	\label{fig:Fig3_bs_trans}
\end{figure}

We achieved a RMSE of approximately 0.3 meV for the E dataset and an RMSE value of approximately 6 meV for the V dataset in this case-study. Figure \ref{fig:Figure2_EV} shows a comparison of predicted versus reference E and V values for the training and test dataset. The $x=y$ line is also included as an indication of the quality of the prediction. A perfect prediction matching reference data lies on this line. It can be seen that the E dataset is trained and predicted with high fidelity. The V dataset also has predictions that mostly lie on or near the $x=y$ line, with some interactions close to zero being predicted away from the line. 

\subsection{$sp3$-orbital strained C (diamond) system}
For a 4 orbital $sp3$ basis used in the C-case study, the DFT Hamiltonian consists of 3 categories of elements.
\begin{enumerate}
\item $E_s, E_p$ - The intra-atomic or on-site energy element for the C $s$ and $p$ orbitals respectively that capture effect of the neighborhood of atoms on a given orbital at an atomic site.
\item $E_{s,p_{x,y,z}}, E_{p_{x},p_{y,z}}$ etc. - Off-diagonal intra-atomic elements that capture interactions between s and p orbitals on the same atom mediated by the potential created by the neighborhood of atoms.
\item $V_{s,s}, V_{s,p_{x,y,z}}, V_{p_{x,y,z},p_{x,y,z}}$ - The inter-atomic Hamiltonian element that captures the strength of the interaction between orbitals situated on different atoms mediated by the potential created by the neighborhood of interacting atoms.
\end{enumerate}
For the multi-orbital basis, the intra-atomic elements are collectively referred to as E, while the inter-atomic elements are collectively referred to as V. We note again that since the two-center approximation does not generally hold in DFT (or in realistically strained systems), we cannot resolve interactions between orbitals in the form of $V_{sp\sigma}, V_{pp\pi}, V_{pp\sigma}$ and other such canonical interactions and we must therefore treat each DFT Hamiltonian element category as a distinct reference output.

\begin{figure}[!htbp]
	\centering
		\includegraphics[width=3in,height=2.5in]{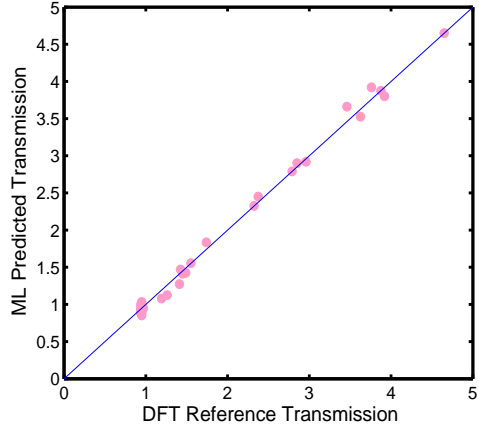}
	\caption{Predicted versus reference transverse-momentum Averaged Transmission calculated for DFT Hamiltonian elements in the test dataset.}
	\label{fig:Fig_Trans}
\end{figure}

To create the reference dataset, we performed self-consistent DFT calculations for a total of 40 strained C systems. The unit cells ranged from single atom FCC unit cells strained arbitrarily to within 4\% of the lattice parameter to larger unitcells oriented along [100], [110] and [111] transport orientations. From the converged DFT calculations, we extract E values for each {atom,orbital-orbital} combination in the calculation and V values for each (atom, orbital, neighbor) combination to form reference output data $y_E$ and $y_V$. This results in a reference/labeled dataset with approxiately 45000 Hamiltonian elements for the C system. A single-zeta numerical basis set consisting of a single $s$ and 3 $p$ orbitals was used as the basis set, with norm-conserving pseudopotentials within the Local Density Approximation (LDA) in DFT \cite{manualatk}. Our observation was that the same $k$-grid density used for Cu was sufficient to ensure converged self-consistent total energy in the C-calculations as well. We use LDA in spite of the well known band-gap underestimation problem for semiconductors because our emphasis in this work is not how well DFT calculations represent the real band structure of C-diamond. Rather, our focus is on how well the method we propose predicts DFT Hamiltonian elements and derived quantities. Being DFT-specifics agnostic, the technique is rather independent of the details of the DFT calculation. As in the case study for strained Cu, one only needs to ensure that a consistent $k$-point grid density and DFT functional is used across all reference calculations so that the calculations are consistent across boundary conditions.

\begin{figure}[!htbp]
	\centering
		\includegraphics[width=\textwidth]{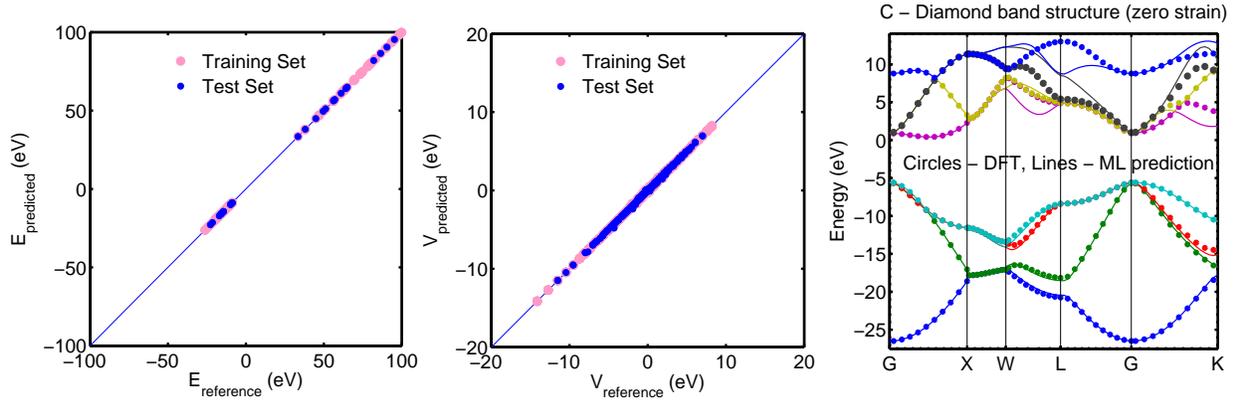}
	\caption{From left to right  - Predicted versus reference intra-atomic interaction energies (left), inter-atomic interaction energies (center) for the $sp3$-orbital C-diamond system studied. The $x=y$ line is included as an indicator for the quality of fit - when predictions match reference/desired data perfectly, they lie on this line. (right) The band structure of unstrained diamond derived from a prediction of the Hamiltonian using this technique.}
	\label{fig:Figure_C}
\end{figure}

As in the case of the Cu-case study, from the converged DFT calculations, we extract E values for each [atom, orbital type] combination in the calculation and V values for each [atom, orbital type, neighbor] combination to form reference output data $y_E$ and $y_V$. The E reference data set for the C case-study consisted of intra-atomic Hamiltonian elements corresponding to a total of 800 atoms in the dataset. Each distinct interaction was represented by a separate column in the reference/output data-matrix. For instance $E_{s,s}, E_{s,p_x}$ and so on are represented by a distinct column in the output data matrix. The feature vector for E prediction for an atom is a row vector consisting of 69 bispectrum coefficients for that atom to which the sum of nearest neighbor direction cosines $\sum{l_i}$, $\sum{m_i}$, $\sum{n_i}$ (where index $i$ is over the nearest neighbors) are appended. The reference feature matrix $X_E$ is thus a $N$$\times$72 matrix where $N$ = 800 is the total number of atoms across all reference calculations for the C-case study. The E output/DFT reference data-matrix is a 800$\times$10 matrix.

The V reference dataset consisted of 44200 Hamiltonian elements. Like in the case of Cu, the V dataset is significantly larger than the E dataset. The feature vector for V prediction consists of 69 bispectrum coefficients for the first and second atoms each divided by the interatomic distance. The 3 direction cosines for the bond corresponding to each {atom, neighbor} combination are appended to this vector followed by the interatomic distance at the end for a total dimensionality of 142 for V features in the dataset. The reference feature matrix $X_v$ is thus a 44200$\times$142 matrix. The output data matrix is a 44200$\times$10 matrix, where each distinct interatomic interaction is captured in a distinct column in the output dataset.

As in the case of the Cu case study, machines for E and V prediction were then trained using the procedure outlined previously. We also use a randomized training:test split of 80:20 percent for E and V training and prediction. Similar to the Cu-case study, we use the RMSE as a measure of goodness-of-fit.
We achieved a RMSE of 1 meV for the E dataset and an RMSE of 18 meV for the V dataset. Figure \ref{fig:Figure2_EV} shows a comparison of predicted versus reference E and V values for the training and test dataset in the C-diamond case-study.

While accurate and computationally efficient prediction of Hamiltonians is a goal of this work, our proposal can be judged as useful only if electronic structure quantities derived from predicted Hamiltonians compare accurately to their \textit{ab initio} DFT calculated counterparts. Since these quantities are not explicitly included in the fitting process, such a comparison also serves as an additional validation of the accuracy of the predicted Hamiltonians. We use band structure and ballistic electronic transmission as a figure-of-merit to compare quantities calculated from predicted Hamiltonians and DFT Hamiltonians computed \textit{ab initio}. For Cu, in the spirit of our previous investigations on the topic \cite{hegde2016lower, hegde2016electron}, we use ballistic transmission as a figure of merit. While ballistic transmission for perfectly periodic structures can be derived from band structure, the applications we cited above make no assumption for periodicity. For C-diamond, which is a semiconductor, band structure is a more appropriate figure of merit. This is because interesting spectral features such as the minimum energy band gap and effective mass, crucial for semiconductors, can be derived directly from band structures.

Figure \ref{fig:Fig3_bs_trans} is an example of such a comparison. The figure shows transverse $k$-resolved ballistic transmission for bulk-unstrained FCC Cu oriented along [100] and for a [110] unit cell uniaxially strained by 2$\%$ along transport direction. We note here that in spite of not having directly included $k$-resolved transmission information in the training process, the DFT-reference and ML-predicted transmission spectra compare quite well qualitatively. A (unitless) transmission RMSE of 0.1 averaged across all k-grid values (21$\times$21 k grid per structure) for each structure in the dataset was achieved. Figure \ref{fig:Fig_Trans} shows a comparison of the transmission averaged over the tranverse Brillouin zone. We computed a transmission RMSE of 0.08 between the reference and predicted \textit{average} transmissions, across a transmission range of 4 (maximum transmission = 4, minimum transmission =0) for the structures studied indicating excellent agreement for the predictions.

The band structure of unstrained C-diamond is predicted and contrasted with its DFT-calculated counterpart in figure \ref{fig:Figure_C}. We note here that in the case of semiconductor band structure modeling using SETB, it is customary to target a limited set of bands that are useful from a conductance perspective. Only those bands that are close to within 25 meV of the Fermi Level are interesting from this point of view. Even so, since the Fermi Level can be modulated by $p$ and $n$ type doping, it is necessary to predict the band structure of the highest valence and lowest conduction band with high degrees of accuracy. We obtained a band structure RMSE of 7 meV for the lowest conduction and highest valence band. If we average across all bands for all test structures we find that our RMSE goes up to 0.1 eV. Data mining our predicted Hamiltonian values reveals that this this arises chiefly due to $p-p$ interactions that are predicted with a higher RMSE of 50 meV (since $p-p$ interactions contribute significantly in bands well above the Fermi Level in undoped C-diamond). A visual inspection of the C-diamond band structure in \ref{fig:Figure_C} confirms a significantly accurate prediction of the valence bands in C while the conduction bands seem to be predicted accurately for some regions in k-space, while they deviate from DFT predictions considerably in other regions.

\section*{Discussion}
The results in previous sections indicate that learning and predicting DFT Hamiltonians is not only feasible, but can also be done with high accuracy of predictions provided meaningful features linking structure to Hamiltonian elements can be created.
It is useful to contrast the method proposed in this paper to other DFT-approximations that exist in the literature in order to understand its relative advantages and limitations. Existing approximations to DFT Hamiltonians can be broadly classified into two main types. In the first, the Hamiltonian elements are computed by fitting a model of intra and inter-atomic interactions to eigenspectra and/or wavefunctions \cite{tan2015tight, tan2016transferable, hegde2014environment}. These techniques create a mathematical map from inputs (atom types, their positions and derived TB Hamiltonian matrix elements) to DFT derived-outputs (wavefunctions, eigenspectra and derived quantities such as band gaps and effective masses). In the second, instead of fitting a model to DFT derived outputs, one alternately maps basis functions and Hamiltonian elements (inputs) to DFT Hamiltonian matrix elements (outputs). If a highly accurate map is found, the derived quantities reflect a similar accuracy. Explicit projection of DFT basis sets and Hamiltonian elements to lower order TB Hamiltonians has been discussed in detail in several publications \cite{urban2011parameterization, elstner1998self, wang2008tight, qian2008quasiatomic}. 

In contrast to the latter set of techniques, we do not form an explicit mathematical map to DFT Hamiltonians. Our method instead interpolates between feature-vectors that map non-linearly to Hamiltonian elements. Instead of defining the map using arbitrary mathematical models that link features to outputs, the use of the kernel-trick in KRR allows defining the map in a high-dimensional feature space without needing to make the map explicit. Since the map changes implicitly for different material systems, we can use the method as is without needing to account for explicit changes in the physics of the material system. In other words, a change in material systems necessitates only re-training of weights for a new dataset that may be generated for this material system. This in contrast, to say, TB, where the use of explicit functional forms relating the parameterized Hamiltonian to its environment requires a continuous assessment of the validity of the model across material systems. Model transferability from one material system to the other is often poor, due to which a wide variety of semi-empirical models exists in the literature (see Goringe et al. \cite{goringe1997tight} for an example of the wide variety of TB models created for different material systems).

No additional assumption regarding the orthogonality of the basis set needs to be made when using Hamiltonian elements predicted by this method. Given that the overlap between numerical basis sets used in real-space DFT codes can completely be defined by doing a dimer calculation for varying radial distance, the overlap integral component of the generalized eigenvalue problem is considered as solved. One can simply pre-compute overlap integrals versus radial distance and use this calculation as a look-up table when needed.

Outside of the decision regarding the best set of physically transparent features to use for the problem at hand, our method obviates the need to fit eigenvalues and other derived features that characterizes almost all semi-empirical approximations to DFT. It is also evident that the formation of the Kernel, training of weights and the selection of hyper-parameters are entirely automated. The creation of training data by can be done by scripting self-consistent DFT calculation runs. Even though other semi-empirical approaches save considerable computational time by avoiding explicit computation to self-consistency, the amount of time taken to develop a particular approximation is never accounted for. In the past, this has been a significant roadblock in our own use of TB approximations to DFT for high-throughput electronic structure screening calculations.

The KRR technique automatically leads to the ability to incorporate multiple outputs/targets within the same procedure by simply adding columns to the output dataset. A new set of weights $\alpha$ for each output of interest is generated in the training process. This is quite convenient since multi-orbital interactions can now be extended very simply in case one decides to incorporate a larger basis set with an increased number of orbitals.
 
Accuracy of the technique in describing specific spectral features can be systematically improved by targeting Hamiltonian elements that govern those spectral features. For instance, in the C-diamond case study, upper conduction bands were not predicted as accurately as valence bands and lower conduction bands. This was attributed to the lower accuracy of the $p-p$ interaction prediction. With the use of weights in the KRR technique, $p-p$ interactions can be targeted for systematic improvement. This, however, comes at the cost of arbitrariness in deciding weights that we prefer avoiding. It appears that alternate ways to systematically improve prediction performance that are grounded in ML theory may be needed for this purpose. For instance, an alternate feature set that leads to a more accurate map for $p-p$ interactions may be needed. Even so, we note that in our experience, generating a model in SETB for arbitrary strain in semiconductors is a venture that results in a development time running into several months at minimum.

Some of the advantages of our method versus alternate DFT-approximation techniques also result in its most important limitation. Creating an explicit model for Hamiltonian elements allows a deeper understanding of the physics of the system, even if the model is not transferable across material systems. For instance, the TB models of Wang et al. \cite{wang1997environment} and Pettifor et al. \cite{pettifor1989new}, relate changes in the atomic environment to Hamiltonian parameters using an intuitive model of how bonding with neighbors affects inter-atomic interactions. Since our technique is non-parametric, this intuitive understanding is lost. In this regard, our opinion is that the opportunity-cost of intuitive understanding when using explicit models is more than compensated for by the ease of implementation, the degree of automation and transferability our method provides.

An alternative route to learning electronic properties directly from structural information has also been adopted in recent literature. For instance, the work of Lopez-Bezanilla \cite{PhysRevB.89.235411} and Sch{\"u}tt et al.\cite{schutt2014represent} shows that the electronic transmission and density of states can be linked to structural information directly. While this approach is immensely powerful and efficient for specific properties, having access to a accurate Hamiltonian prediction enables the computation of all of all such derived quantities simultaneously instead of having to train multiple machines on multiple datasets with different reference electronic properties.

The advantage of supervised machine learning techniques in making sense of existing, labeled data is also an important limitation in some sense. It is clear from the KRR formalism presented above that the technique is interpolation-based. In other words, the quality of the prediction is dependent on having data representations in the training data set that are somewhat similar to the candidate for prediction. In certain cases, this is a limitation that can be ignored - for instance, the space of known and achievable strains in Si used in logic technology lead to a a rather small number of interesting cases of strained Si within the space of all possible strains and deformations of FCC-Si. In such a case it is very easy to do a dense sampling of strained Si candidates, so that predictions on this system can be interpolated with high accuracy. Predictions of the band structure of silicene from a data set trained on strained FCC-Si, however, are not feasible in KRR (and possibly in other ML techniques) due to their supervised nature, where some labeled data is required in advance of the prediction. This necessitates enlarging the training dataset to incorporates at least some sense of the physical interactions that may be encountered in a target application.

Extensions of the technique to multi-component/multi-element systems involves two distinct steps. The first is the computation of a bispectrum/alternate-feature vector for a multi-component system. The second is extending the number of interactions that need to be mapped from DFT using machine learning. As a case in point, we consider the semiconductor InAs. Instead of a single $s-p$ interaction we included in the output reference dataset, now, two such interactions need to be included, since the $s-p$ interactions involving an In $s$ orbital and an As $p$ orbital is quite different that those involving As $s$ and In $p$ orbitals. Apart from this modification, no additional changes are needed in the technique when applying it to multi-component systems.
   
In conclusion, we have proposed a technique based on Machine Learning for automated learning and prediction of DFT Hamiltonians. The method is based on a non-parametric, non-linear mapping of atomic bispectrum features to DFT Hamiltonian elements using Kernel Ridge Regression. It is transferable across material systems and is independent of the specifics of the DFT functional used. We have demonstrated that the method can provide accurate prediction of DFT Hamiltonians and that the derived electronic structure quantities compare accurately to their DFT calculated counterparts. We believe that the method provides a promising starting point for highly-automated learning and prediction of DFT Hamiltonian elements towards application in high-throughput electronic structure calculations.


\begin{thebibliography}{10}
\expandafter\ifx\csname url\endcsname\relax
  \def\url#1{\texttt{#1}}\fi
\expandafter\ifx\csname urlprefix\endcsname\relax\def\urlprefix{URL }\fi
\providecommand{\bibinfo}[2]{#2}
\providecommand{\eprint}[2][]{\url{#2}}

\bibitem{bowler2012methods}
\bibinfo{author}{Bowler, D.} \& \bibinfo{author}{Miyazaki, T.}
\newblock \bibinfo{title}{Methods in electronic structure calculations}.
\newblock \emph{\bibinfo{journal}{Reports on Progress in Physics}}
  \textbf{\bibinfo{volume}{75}}, \bibinfo{pages}{036503}
  (\bibinfo{year}{2012}).

\bibitem{goringe1997tight}
\bibinfo{author}{Goringe, C.}, \bibinfo{author}{Bowler, D.} \emph{et~al.}
\newblock \bibinfo{title}{Tight-binding modelling of materials}.
\newblock \emph{\bibinfo{journal}{Reports on Progress in Physics}}
  \textbf{\bibinfo{volume}{60}}, \bibinfo{pages}{1447} (\bibinfo{year}{1997}).

\bibitem{klimeck2000si}
\bibinfo{author}{Klimeck, G.} \emph{et~al.}
\newblock \bibinfo{title}{Si tight-binding parameters from genetic algorithm
  fitting}.
\newblock \emph{\bibinfo{journal}{Superlattices and Microstructures}}
  \textbf{\bibinfo{volume}{27}}, \bibinfo{pages}{77--88}
  (\bibinfo{year}{2000}).

\bibitem{hegde2014environment}
\bibinfo{author}{Hegde, G.}, \bibinfo{author}{Povolotskyi, M.},
  \bibinfo{author}{Kubis, T.}, \bibinfo{author}{Boykin, T.} \&
  \bibinfo{author}{Klimeck, G.}
\newblock \bibinfo{title}{An environment-dependent semi-empirical tight binding
  model suitable for electron transport in bulk metals, metal alloys, metallic
  interfaces, and metallic nanostructures. i. model and validation}.
\newblock \emph{\bibinfo{journal}{Journal of Applied Physics}}
  \textbf{\bibinfo{volume}{115}} (\bibinfo{year}{2014}).

\bibitem{friedman2001elements}
\bibinfo{author}{Friedman, J.}, \bibinfo{author}{Hastie, T.} \&
  \bibinfo{author}{Tibshirani, R.}
\newblock \emph{\bibinfo{title}{The elements of statistical learning}},
  vol.~\bibinfo{volume}{1} (\bibinfo{publisher}{Springer series in statistics
  Springer, Berlin}, \bibinfo{year}{2001}).

\bibitem{bartok2010gaussian}
\bibinfo{author}{Bart{\'o}k, A.~P.}, \bibinfo{author}{Payne, M.~C.},
  \bibinfo{author}{Kondor, R.} \& \bibinfo{author}{Cs{\'a}nyi, G.}
\newblock \bibinfo{title}{Gaussian approximation potentials: The accuracy of
  quantum mechanics, without the electrons}.
\newblock \emph{\bibinfo{journal}{Physical review letters}}
  \textbf{\bibinfo{volume}{104}}, \bibinfo{pages}{136403}
  (\bibinfo{year}{2010}).

\bibitem{rupp2012fast}
\bibinfo{author}{Rupp, M.}, \bibinfo{author}{Tkatchenko, A.},
  \bibinfo{author}{M{\"u}ller, K.-R.} \& \bibinfo{author}{Von~Lilienfeld,
  O.~A.}
\newblock \bibinfo{title}{Fast and accurate modeling of molecular atomization
  energies with machine learning}.
\newblock \emph{\bibinfo{journal}{Physical review letters}}
  \textbf{\bibinfo{volume}{108}}, \bibinfo{pages}{058301}
  (\bibinfo{year}{2012}).

\bibitem{behler2007generalized}
\bibinfo{author}{Behler, J.} \& \bibinfo{author}{Parrinello, M.}
\newblock \bibinfo{title}{Generalized neural-network representation of
  high-dimensional potential-energy surfaces}.
\newblock \emph{\bibinfo{journal}{Physical review letters}}
  \textbf{\bibinfo{volume}{98}}, \bibinfo{pages}{146401}
  (\bibinfo{year}{2007}).

\bibitem{snyder2012finding}
\bibinfo{author}{Snyder, J.~C.}, \bibinfo{author}{Rupp, M.},
  \bibinfo{author}{Hansen, K.}, \bibinfo{author}{M{\"u}ller, K.-R.} \&
  \bibinfo{author}{Burke, K.}
\newblock \bibinfo{title}{Finding density functionals with machine learning}.
\newblock \emph{\bibinfo{journal}{Physical review letters}}
  \textbf{\bibinfo{volume}{108}}, \bibinfo{pages}{253002}
  (\bibinfo{year}{2012}).

\bibitem{schutt2014represent}
\bibinfo{author}{Sch{\"u}tt, K.} \emph{et~al.}
\newblock \bibinfo{title}{How to represent crystal structures for machine
  learning: Towards fast prediction of electronic properties}.
\newblock \emph{\bibinfo{journal}{Physical Review B}}
  \textbf{\bibinfo{volume}{89}}, \bibinfo{pages}{205118}
  (\bibinfo{year}{2014}).

\bibitem{montavon2012learning}
\bibinfo{author}{Montavon, G.} \emph{et~al.}
\newblock \bibinfo{title}{Learning invariant representations of molecules for
  atomization energy prediction}.
\newblock In \emph{\bibinfo{booktitle}{Advances in Neural Information
  Processing Systems}}, \bibinfo{pages}{440--448} (\bibinfo{year}{2012}).

\bibitem{hansen2015machine}
\bibinfo{author}{Hansen, K.} \emph{et~al.}
\newblock \bibinfo{title}{Machine learning predictions of molecular properties:
  Accurate many-body potentials and nonlocality in chemical space}.
\newblock \emph{\bibinfo{journal}{The journal of physical chemistry letters}}
  \textbf{\bibinfo{volume}{6}}, \bibinfo{pages}{2326--2331}
  (\bibinfo{year}{2015}).

\bibitem{von2015fourier}
\bibinfo{author}{von Lilienfeld, O.~A.}, \bibinfo{author}{Ramakrishnan, R.},
  \bibinfo{author}{Rupp, M.} \& \bibinfo{author}{Knoll, A.}
\newblock \bibinfo{title}{Fourier series of atomic radial distribution
  functions: A molecular fingerprint for machine learning models of quantum
  chemical properties}.
\newblock \emph{\bibinfo{journal}{International Journal of Quantum Chemistry}}
  \textbf{\bibinfo{volume}{115}}, \bibinfo{pages}{1084--1093}
  (\bibinfo{year}{2015}).

\bibitem{bartok2013representing}
\bibinfo{author}{Bart{\'o}k, A.~P.}, \bibinfo{author}{Kondor, R.} \&
  \bibinfo{author}{Cs{\'a}nyi, G.}
\newblock \bibinfo{title}{On representing chemical environments}.
\newblock \emph{\bibinfo{journal}{Physical Review B}}
  \textbf{\bibinfo{volume}{87}}, \bibinfo{pages}{184115}
  (\bibinfo{year}{2013}).

\bibitem{thompson2015spectral}
\bibinfo{author}{Thompson, A.~P.}, \bibinfo{author}{Swiler, L.~P.},
  \bibinfo{author}{Trott, C.~R.}, \bibinfo{author}{Foiles, S.~M.} \&
  \bibinfo{author}{Tucker, G.~J.}
\newblock \bibinfo{title}{Spectral neighbor analysis method for automated
  generation of quantum-accurate interatomic potentials}.
\newblock \emph{\bibinfo{journal}{Journal of Computational Physics}}
  \textbf{\bibinfo{volume}{285}}, \bibinfo{pages}{316--330}
  (\bibinfo{year}{2015}).

\bibitem{bartok2015gaussian}
\bibinfo{author}{Bart{\'o}k, A.~P.} \& \bibinfo{author}{Cs{\'a}nyi, G.}
\newblock \bibinfo{title}{Gaussian approximation potentials: A brief tutorial
  introduction}.
\newblock \emph{\bibinfo{journal}{International Journal of Quantum Chemistry}}
  \textbf{\bibinfo{volume}{115}}, \bibinfo{pages}{1051--1057}
  (\bibinfo{year}{2015}).

\bibitem{podolskiy2004compact}
\bibinfo{author}{Podolskiy, A.} \& \bibinfo{author}{Vogl, P.}
\newblock \bibinfo{title}{Compact expression for the angular dependence of
  tight-binding hamiltonian matrix elements}.
\newblock \emph{\bibinfo{journal}{Physical Review B}}
  \textbf{\bibinfo{volume}{69}}, \bibinfo{pages}{233101}
  (\bibinfo{year}{2004}).

\bibitem{rupp2015machine}
\bibinfo{author}{Rupp, M.}
\newblock \bibinfo{title}{Machine learning for quantum mechanics in a
  nutshell}.
\newblock \emph{\bibinfo{journal}{International Journal of Quantum Chemistry}}
  \textbf{\bibinfo{volume}{115}}, \bibinfo{pages}{1058--1073}
  (\bibinfo{year}{2015}).

\bibitem{rasmussen2006gaussian}
\bibinfo{author}{Rasmussen, C.~E.}
\newblock \bibinfo{title}{Gaussian processes for machine learning}
  (\bibinfo{year}{2006}).

\bibitem{smola2004tutorial}
\bibinfo{author}{Smola, A.~J.} \& \bibinfo{author}{Sch{\"o}lkopf, B.}
\newblock \bibinfo{title}{A tutorial on support vector regression}.
\newblock \emph{\bibinfo{journal}{Statistics and computing}}
  \textbf{\bibinfo{volume}{14}}, \bibinfo{pages}{199--222}
  (\bibinfo{year}{2004}).

\bibitem{abu2012learning}
\bibinfo{author}{Abu-Mostafa, Y.~S.}, \bibinfo{author}{Magdon-Ismail, M.} \&
  \bibinfo{author}{Lin, H.-T.}
\newblock \emph{\bibinfo{title}{Learning from data}}, vol.~\bibinfo{volume}{4}
  (\bibinfo{publisher}{AMLBook Singapore}, \bibinfo{year}{2012}).

\bibitem{hegde2015feasibility}
\bibinfo{author}{Hegde, G.} \& \bibinfo{author}{Bowen, R.~C.}
\newblock \bibinfo{title}{On the feasibility of ab initio electronic structure
  calculations for cu using a single s orbital basis}.
\newblock \emph{\bibinfo{journal}{AIP Advances}} \textbf{\bibinfo{volume}{5}},
  \bibinfo{pages}{107142} (\bibinfo{year}{2015}).

\bibitem{manualatk}
\bibinfo{author}{Manual, A.~T.}
\newblock \bibinfo{title}{Atk version 2015.1}.
\newblock \emph{\bibinfo{journal}{QuantumWise A/S (www. quantumwise. com)}} .

\bibitem{hegde2016lower}
\bibinfo{author}{Hegde, G.}, \bibinfo{author}{Bowen, R.} \&
  \bibinfo{author}{Rodder, M.~S.}
\newblock \bibinfo{title}{Lower limits of line resistance in nanocrystalline
  back end of line cu interconnects}.
\newblock \emph{\bibinfo{journal}{Applied Physics Letters}}
  \textbf{\bibinfo{volume}{109}} (\bibinfo{year}{2016}).

\bibitem{hegde2016electron}
\bibinfo{author}{Hegde, G.}, \bibinfo{author}{Bowen, R.~C.} \&
  \bibinfo{author}{Rodder, M.~S.}
\newblock \bibinfo{title}{Is electron transport in nanocrystalline cu
  interconnects surface dominated or grain boundary dominated?}
\newblock In \emph{\bibinfo{booktitle}{Interconnect Technology
  Conference/Advanced Metallization Conference (IITC/AMC), 2016 IEEE
  International}}, \bibinfo{pages}{114--116} (\bibinfo{organization}{IEEE},
  \bibinfo{year}{2016}).

\bibitem{tan2015tight}
\bibinfo{author}{Tan, Y.~P.}, \bibinfo{author}{Povolotskyi, M.},
  \bibinfo{author}{Kubis, T.}, \bibinfo{author}{Boykin, T.~B.} \&
  \bibinfo{author}{Klimeck, G.}
\newblock \bibinfo{title}{Tight-binding analysis of si and gaas ultrathin
  bodies with subatomic wave-function resolution}.
\newblock \emph{\bibinfo{journal}{Physical Review B}}
  \textbf{\bibinfo{volume}{92}}, \bibinfo{pages}{085301}
  (\bibinfo{year}{2015}).

\bibitem{tan2016transferable}
\bibinfo{author}{Tan, Y.}, \bibinfo{author}{Povolotskyi, M.},
  \bibinfo{author}{Kubis, T.}, \bibinfo{author}{Boykin, T.~B.} \&
  \bibinfo{author}{Klimeck, G.}
\newblock \bibinfo{title}{Transferable tight-binding model for strained group
  iv and iii-v materials and heterostructures}.
\newblock \emph{\bibinfo{journal}{Physical Review B}}
  \textbf{\bibinfo{volume}{94}}, \bibinfo{pages}{045311}
  (\bibinfo{year}{2016}).

\bibitem{urban2011parameterization}
\bibinfo{author}{Urban, A.}, \bibinfo{author}{Reese, M.},
  \bibinfo{author}{Mrovec, M.}, \bibinfo{author}{Els{\"a}sser, C.} \&
  \bibinfo{author}{Meyer, B.}
\newblock \bibinfo{title}{Parameterization of tight-binding models from density
  functional theory calculations}.
\newblock \emph{\bibinfo{journal}{Physical Review B}}
  \textbf{\bibinfo{volume}{84}}, \bibinfo{pages}{155119}
  (\bibinfo{year}{2011}).

\bibitem{elstner1998self}
\bibinfo{author}{Elstner, M.} \emph{et~al.}
\newblock \bibinfo{title}{Self-consistent-charge density-functional
  tight-binding method for simulations of complex materials properties}.
\newblock \emph{\bibinfo{journal}{Physical Review B}}
  \textbf{\bibinfo{volume}{58}}, \bibinfo{pages}{7260} (\bibinfo{year}{1998}).

\bibitem{wang2008tight}
\bibinfo{author}{Wang, C.-Z.} \emph{et~al.}
\newblock \bibinfo{title}{Tight-binding hamiltonian from first-principles
  calculations}.
\newblock In \emph{\bibinfo{booktitle}{Scientific Modeling and Simulations}},
  \bibinfo{pages}{81--95} (\bibinfo{publisher}{Springer},
  \bibinfo{year}{2008}).

\bibitem{qian2008quasiatomic}
\bibinfo{author}{Qian, X.} \emph{et~al.}
\newblock \bibinfo{title}{Quasiatomic orbitals for ab initio tight-binding
  analysis}.
\newblock \emph{\bibinfo{journal}{Physical Review B}}
  \textbf{\bibinfo{volume}{78}}, \bibinfo{pages}{245112}
  (\bibinfo{year}{2008}).

\bibitem{wang1997environment}
\bibinfo{author}{Wang, C.} \emph{et~al.}
\newblock \bibinfo{title}{Environment-dependent tight-binding potential model}.
\newblock In \emph{\bibinfo{booktitle}{MRS Proceedings}}, vol.
  \bibinfo{volume}{491}, \bibinfo{pages}{211} (\bibinfo{organization}{Cambridge
  Univ Press}, \bibinfo{year}{1997}).

\bibitem{pettifor1989new}
\bibinfo{author}{Pettifor, D.}
\newblock \bibinfo{title}{New many-body potential for the bond order}.
\newblock \emph{\bibinfo{journal}{Physical review letters}}
  \textbf{\bibinfo{volume}{63}}, \bibinfo{pages}{2480} (\bibinfo{year}{1989}).

\bibitem{PhysRevB.89.235411}
\bibinfo{author}{Lopez-Bezanilla, A.} \& \bibinfo{author}{von Lilienfeld,
  O.~A.}
\newblock \bibinfo{title}{Modeling electronic quantum transport with machine
  learning}.
\newblock \emph{\bibinfo{journal}{Phys. Rev. B}} \textbf{\bibinfo{volume}{89}},
  \bibinfo{pages}{235411} (\bibinfo{year}{2014}).
\newblock \urlprefix\url{http://link.aps.org/doi/10.1103/PhysRevB.89.235411}.

\end{thebibliography}


\section*{Acknowledgements}

We acknowledge useful discussions with Borna Obradovic and Ryan Hatcher. We thank Mark Rodder for a thorough reading of the manuscript and for helpful suggestions. 

\section*{Author contributions statement}

G.H. devised the method, and performed the calculations. G.H. performed the analysis. G.H. and R.C.B wrote the manuscript. All authors reviewed the manuscript. 

\section*{Additional information}

\textbf{Competing financial interests} (mandatory statement). 
The authors declare no competing financial interests.


%

\end{document}